\documentclass[5p,times]{elsarticle}

\usepackage[utf8]{inputenc}
\usepackage{amssymb}
\usepackage{amsmath}
\usepackage{graphicx}
\usepackage{placeins}
\usepackage{pgf}
\usepackage{siunitx}
\DeclareSIUnit[number-unit-product = {}]\c{c}
\DeclareSIUnit[number-unit-product = {}]\bit{bit}
\DeclareSIUnit[number-unit-product = {}]\byte{byte}
\DeclareSIUnit[number-unit-product = {}]\muon{\ensuremath{\mu}}

\usepackage{textcomp}
\usepackage{xspace}

\newcommand{\mupixS}{\textsc{MuPix7}\xspace}

\newcommand{\mupix}{\textsc{MuPix}\xspace}

\begin{document}
\begin{frontmatter}
\journal{Nuclear Instruments and Methods A}
\title{Efficiency and timing performance of the \mupixS high-voltage monolithic active
pixel sensor}
\author[A]{Heiko Augustin}
\author[B]{Niklaus Berger\corref{mycorrespondingauthor}}
\ead{niberger@uni-mainz.de}
\author[A]{Sebastian Dittmeier}
\author[A,B]{Carsten Grzesik}
\author[A]{Jan Hammerich}
\author[B]{Ulrich Hartenstein}
\author[B]{Qinhua Huang\fnref{ParisIIfootnote}}
\author[A]{Lennart Huth}
\author[A]{David Maximilian Immig}
\author[A]{Moritz Kiehn\fnref{myotherfootnote}}
\author[B]{Alexandr Kozlinskiy}
\author[A]{Frank Meier Aeschbacher}
\author[A]{Annie Meneses Gonz\'alez}
\author[C]{Ivan Peri\'c}
\author[A]{Ann-Kathrin Perrevoort}
\author[A]{Andr\'e Sch\"oning}
\author[A]{Shruti Shrestha\fnref{myfootnote}}
\author[B]{Dorothea vom Bruch\fnref{Parisfootnote}}
\author[B]{Frederik Wauters}
\author[A]{Dirk Wiedner}
\address[A]{Physikalisches Institut, Heidelberg University, Heidelberg, Germany}
\fntext[ParisIIfootnote]{Now at \'Ecole Polytechnique, CNRS/IN2P3, Palaiseau, France}
\fntext[myotherfootnote]{Now at D\'epartement de physique nucl\'eaire et corpusculaire, University of Geneva, Switzerland}
\fntext[myfootnote]{Now at Middle Tennessee State University, Murfreesboro, Tennessee, USA}
\fntext[Parisfootnote]{Now at LPNHE, Sorbonne Universit\'e, Universit\'e Paris Diderot, CNRS/IN2P3, Paris, France}
\address[B]{Institut f\"ur Kernphysik and PRISMA cluster of excellence, Mainz University, Mainz, Germany}
\address[C]{Institut f\"ur Prozessdatenverarbeitung und Elektronik, KIT, Karlsruhe, Germany}
\cortext[mycorrespondingauthor]{Corresponding author}

\begin{abstract}
The \mupixS is a prototype high voltage monolithic active pixel sensor with
$103 \times \SI{80}{\micro\meter\squared}$ pixels thinned to \SI{64}{\micro\meter} and 
incorporating the complete read-out circuitry including a \SI{1.25}{Gbit/s}
differential data link. Using data taken at the DESY electron test beam, we
demonstrate an efficiency of \SI{99.3}{\percent} and a time resolution of 
\SI{14}{\nano\second}. The efficiency and time resolution are studied with
sub-pixel resolution and reproduced in simulations.
\end{abstract}
\begin{keyword}
  Silicon Pixel Detectors \sep
  Monolithic Sensors 
  
\end{keyword}
\end{frontmatter}


\section{Motivation}

High-rate precision experiments in particle physics require fast, high-resolution
tracking detectors.
For low momentum particles, multiple Coulomb scattering in detector material is
the main source of tracking uncertainty and thus the material in the tracking
volume has to be minimized.
High voltage monolithic active pixel sensors (HV-MAPS) 
\cite{Peric:2007zz,Peric2010,Peric2010504,Peric:2012bp,Peric:2013cka} allow for the
construction of fast, pixelated and thin particle detectors. A commercial 
high voltage process (AMS \SI{180}{nm} HV-CMOS) allows for deep $n$-wells in a
$p$-doped substrate; a high voltage of around \SI{90}{V} between the $n$-wells
and the substrate creates a thin depletion zone with very large fields, leading
to fast charge collection.
Inside of the deep $n$-well, shallow $p$ and $n$-wells can be placed, allowing
for the implementation of CMOS transistors; in particular, an amplifier circuit
can be placed directly inside the pixel. 
As the active region is very thin, most of the substrate can be removed after
manufacturing and sensors with a thickness down to only \SI{50}{\micro m} can be
obtained.

The Mu3e experiment \cite{mu3e-rp} searches for the lepton-flavour violating 
decay $\mu^+ \rightarrow e^+e^-e^+$ aiming for a sensitivity of one in \num{1e16}
decays. 
In order to track more than \num{1e9} electrons and positrons per second and
determine their momentum with high precision, four layers of HV-MAPS sensors
in a barrel geometry will be used \cite{BergerEtAl:2013:tme}. 
For accidental background suppression and to ease the reconstruction of tracks
in this very high rate environment, a good time resolution of the tracking 
detector is essential.

In preparation for the Mu3e experiment, we have produced and characterized a 
series of HV-MAPS prototypes, the \mupix chips \cite{Augustin:2015mqa}.
In this series, the \mupixS, is the first prototype
incorporating the complete functionality required for the experiment:
signals are amplified in the pixel and driven to the chip periphery by a 
source-follower.
A comparator compares the signal to a threshold, which is adjustable individually
for each pixel.
In case a hit is detected, an eight-bit Gray-encoded timestamp is stored.
Priority logic coupled to a state machine collects and serializes the hit times
and column and row addresses and
sends them off-chip using a \SI{1.25}{Gbit/s} low-voltage differential signaling 
readout link\footnote{the chip and link are 
fully functional up to \SI{1.6}{Gbit/s},
we however typically operate it at the lower frequency} with 8bit/10bit encoding \cite{Widmer1983}.
The active area of the chip is \SI{3.2}{mm} $\times$ \SI{3.2}{mm} incorporating a 
matrix of 32 $\times$ 40 pixels resulting in a pitch of \SI{103}{\micro m} in column direction 
and \SI{80}{\micro m} in row direction.
The sensors were thinned to \SI{64}{\micro m}.
A detailed description of the system and global efficiency and noise measurements 
can be found in reference \cite{Augustin:2016hzx}.

\begin{figure}
	\centering
		\includegraphics[width=0.49\textwidth]{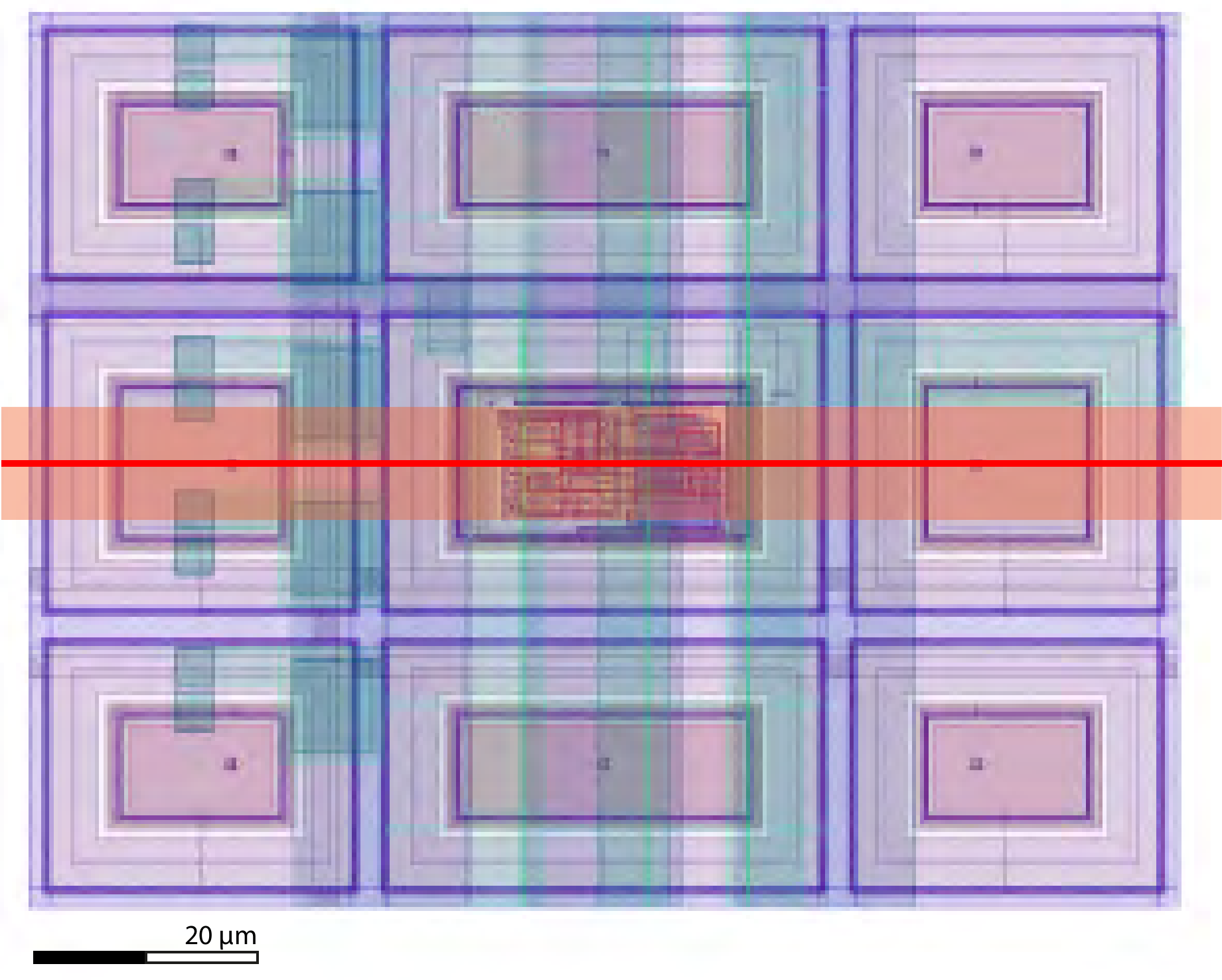}
	\caption{Design view of the MuPix7 pixel unit cell; the nine deep n-wells serving as charge collection electrodes are clearly visible. The central one contains the amplifier and line driver.
	The 2-D TCAD simulations (see section~\ref{sec:Simulation}) were performed for
	the geometry along the central red line and are compared to data from the
	orange shaded area.}
	\label{fig:MUPIX7_Sensor}
\end{figure}

In this work, we present a detailed study of the efficiency and time resolution
of the \mupixS with sub-pixel spatial resolution.
This is of particular interest, as there are nine separate, shorted 
charge-collecting diodes in each pixel, see figure~\ref{fig:MUPIX7_Sensor}, and it 
has to be ensured that charge is also collected efficiently from the space 
between the diodes. In fact, charge sharing between pixels is known to be the 
largest contribution to efficiency losses in \mupix chips. Optimizing the diode
geometry to ensure efficient charge collection with the lowest possible 
capacitance is one of the main goals for further detector development and we
thus aim to verify technology computer aided design (TCAD) simulations of 
our geometry with beam measurements.

The timing resolution in our binary, single threshold readout scheme is mainly
driven by variations in signal size (time-walk).
In the present study, we try to also identify contributions from charge drift 
and diffusion as well as signal transmission times.

The homogeneity of the efficiency and timing resolution over the whole chip is 
of great importance in view of the next step in HV-MAPS development, the 
production of a large \SI{10}{mm} $\times$ \SI{20}{mm} sensor. 

\section{Measurement set-up}

We present results from data taken in March 2016 at the DESY-II beam test 
facility \cite{DESYTB}. 
Photons are produced via bremsstrahlung by placing carbon fibres in 
the electron beam of the DESY-II synchrotron; they subsequently convert into 
electron-positron pairs via pair-production in a metal target. 
With a dipole magnet, the electrons and positrons are separated and electrons 
with an energy of \SI{4}{GeV} were selected with a collimator, resulting in a
rate of approximately \SI{3}{kHz}.  

Reference tracks were obtained from the EUDET Telescope Duranta \cite{Jansen2016}, 
which consists of 
six planes of monolithic active pixel sensors (MIMOSA 26, \cite{HuGuo:2010zz}). 
They have a pixel size of \SI{18.4}{\micro m} $\times$
\SI{18.4}{\micro m} and an active area of \SI{224}{mm\squared}. 
The sensors are \SI{50}{\micro m} thick and glued on a \SI{50}{\micro m}
protective foil, which sums up to \SI{0.7}{permil} of a radiation length. 
The maximum track rate that can be read out is $\sim$ \SI{100}{kHz}. 
After the first three tracking planes, a device under test (DUT) can be placed 
on a rotational stage. 
In this way, a \mupixS chip was placed as DUT in the center of the telescope. 
In addition to the pixel sensors, the telescope is equipped with four 
scintillators -- two before, and two after the pixel sensors -- 
read out by photomultiplier tubes (PMTs) used as triggers. 
A trigger logic unit and the EUDAQ data acquisition 
framework \cite{Perrey:2014tda} were used to combine the data streams from the 
telescope, the scintillators and the DUT.
The \mupixS DUT was operated with a frequency for the timestamps of 
\SI{62.5}{MHz}.  
We directly fed the scintillator coincidence signal 
into the field programmable gate array (FPGA) used to read out the \mupixS 
sensor, sampling with a clock frequency of \SI{500}{MHz}, therefore storing the 
most precise timing information available.
The telescope planes were operated with threshold levels corresponding to a
collected charge in each individual pixel of at least five times the RMS noise. 

A track based alignment of the telescope planes was performed using the
Millepede~II algorithm with the general broken line track model 
\cite{BlobelEtAl:2011:fac}. This resulted in 
mean residuals on the telescope planes in a range from \SI{0}{\micro m} to
\SI{\pm 1.5}{\micro m} and residual widths below \SI{\pm 6}{\micro m}. 

\section{Efficiency measurement}

\begin{figure}[tb!]
  \centering
  \includegraphics[width=0.49\textwidth]{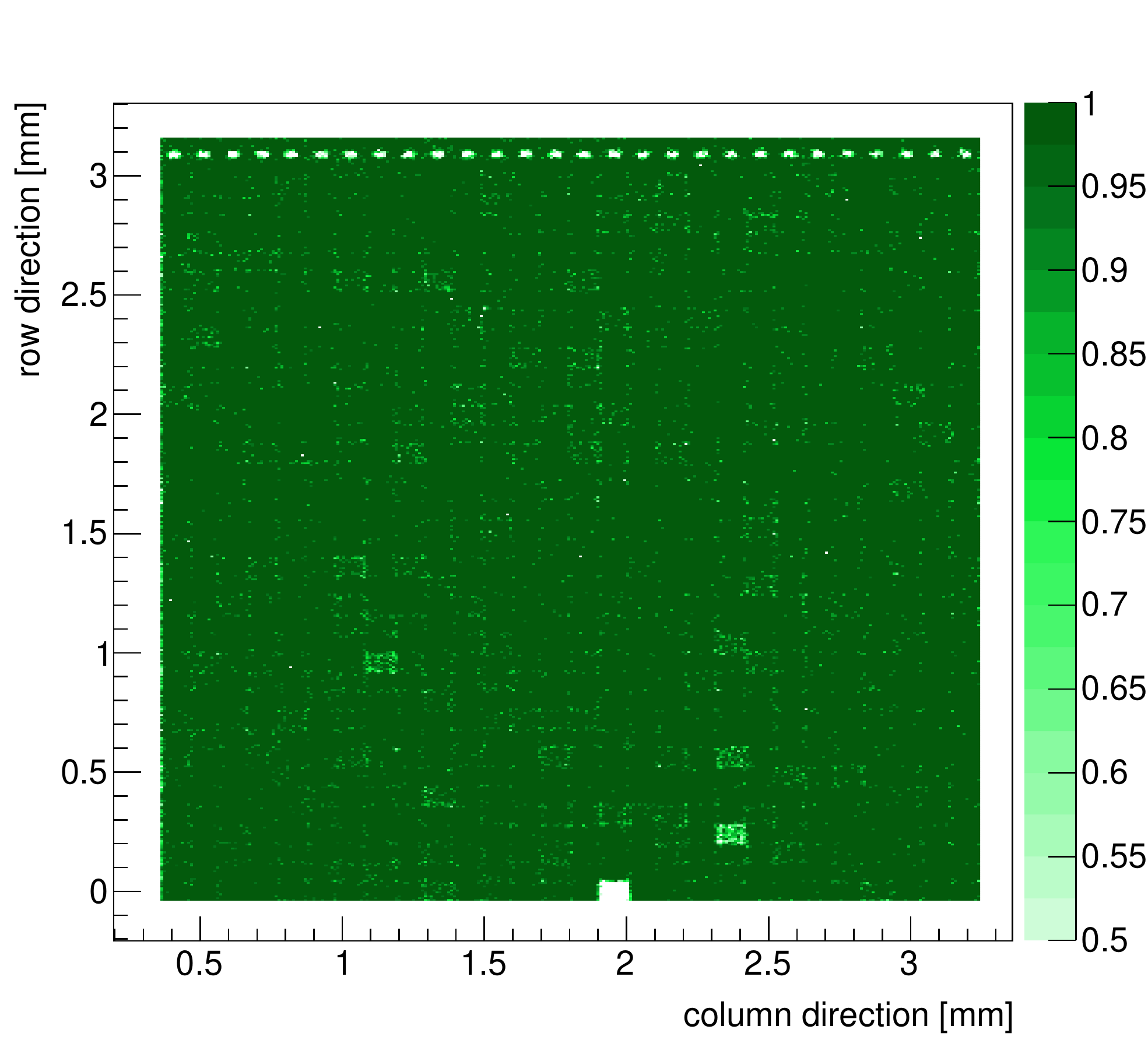}
  \caption{Efficiency map, measured with HV=\SI{-85}{V} and a threshold of \SI{65}{mV}. 
	Efficiency drops in the corners of the pixels, where charge is shared between
	four pixels are visible. In the topmost row, one of the nine diodes in each
	pixel was not connected for test purposes, leading to a pattern of inefficiencies.
	One pixel in the bottommost row was so noisy that it had to be removed from the
	analysis.}
  \label{fig:efficency_map_85V}
\end{figure}

\begin{figure}[tb!]
  \centering
  \includegraphics[width=0.49\textwidth]{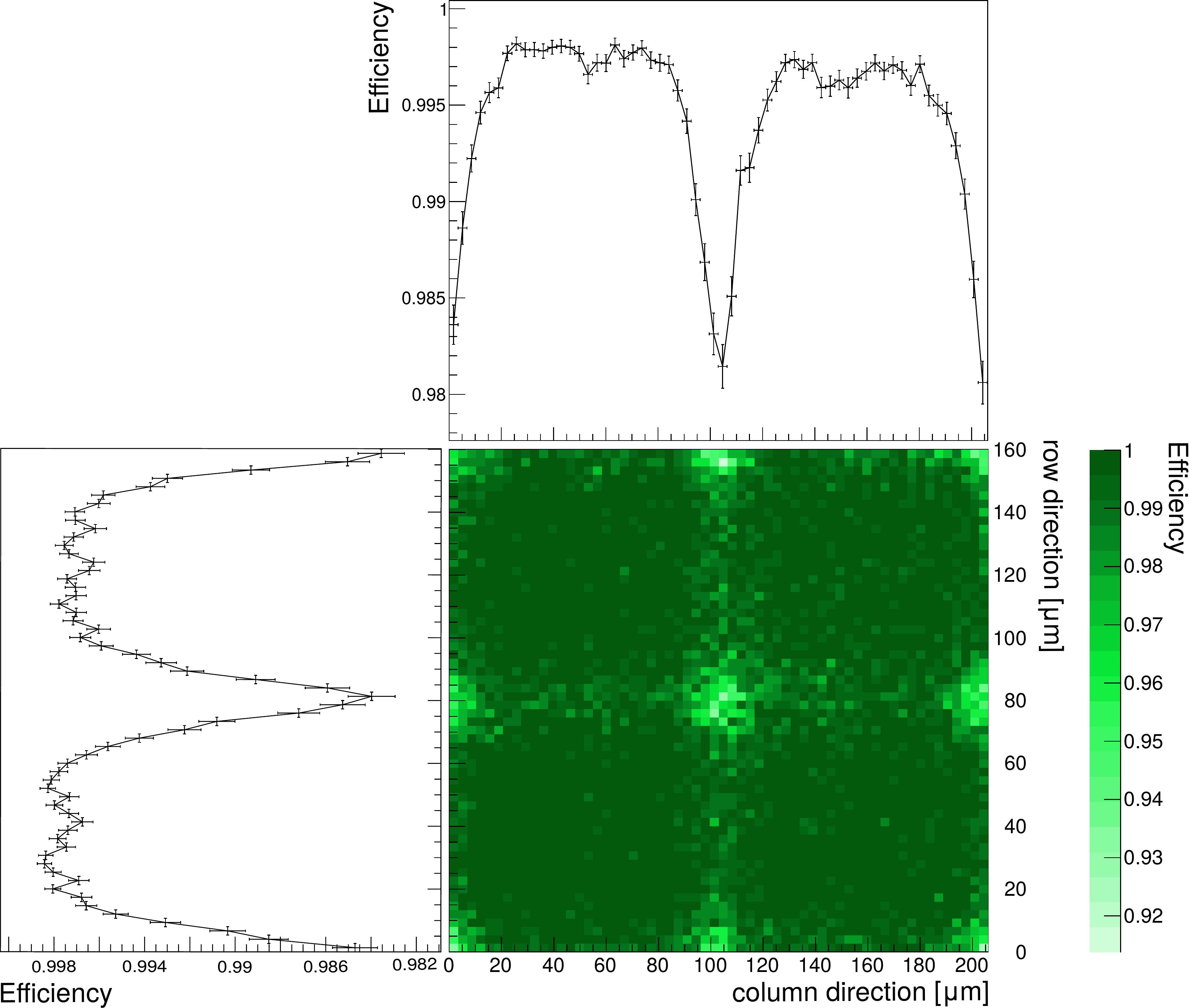}
  \caption{Efficiency measured with HV=\SI{-85}{V} and a threshold of \SI{65}{mV}. 
	Sub-matrices of 2$\times$2 pixels are stacked on top of each other, excluding the outer two columns / rows.  }
  \label{fig:efficency_folded_85V}
\end{figure}



\begin{figure}[tb!]
  \centering
  \includegraphics[width=0.49\textwidth]{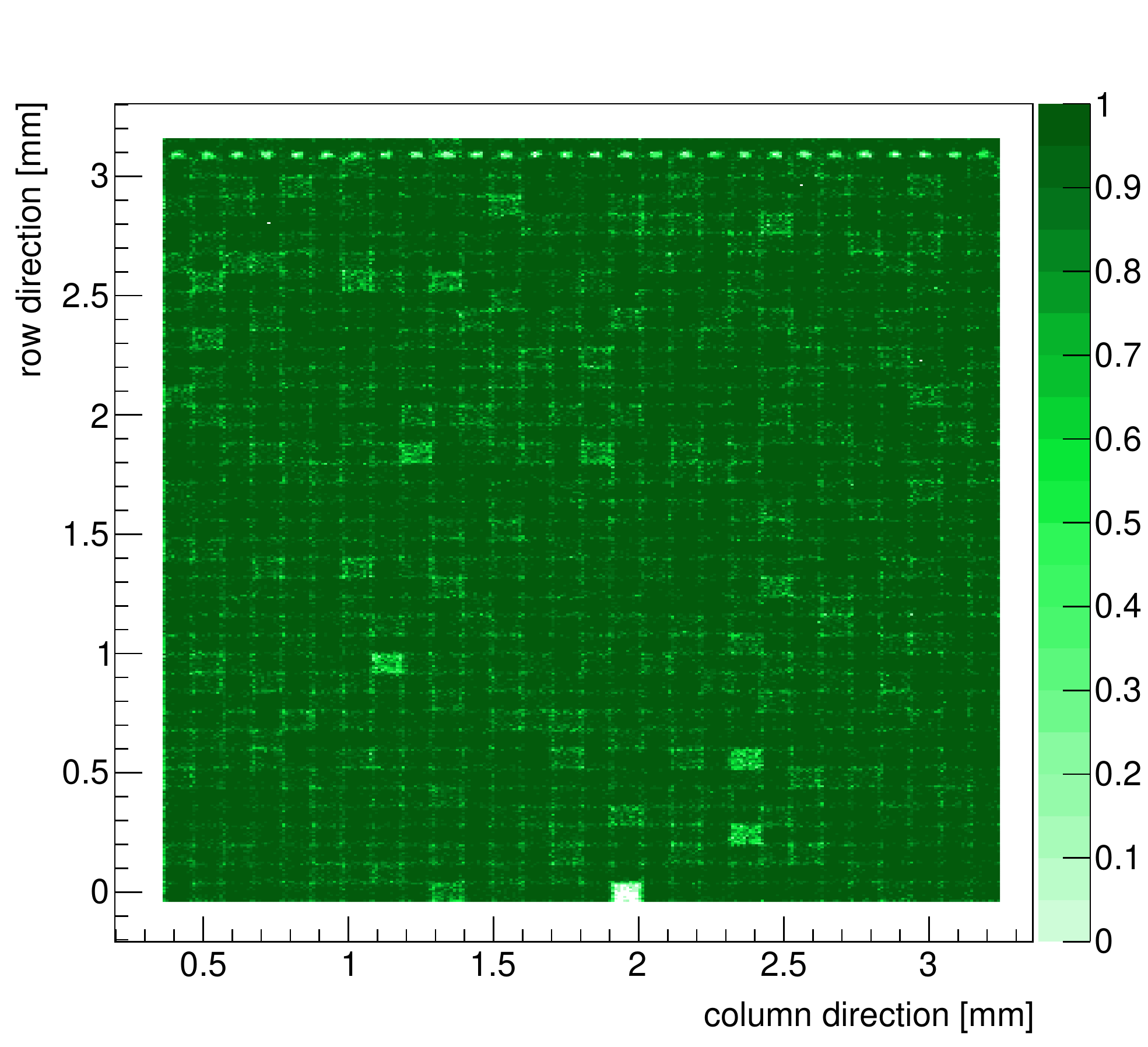}
  \caption{Efficiency map, measured with HV=\SI{-40}{V} and a threshold of \SI{70}{mV}.}
  \label{fig:efficency_map_40V}
\end{figure}

\begin{figure}[tb!]
  \centering
  \includegraphics[width=0.49\textwidth]{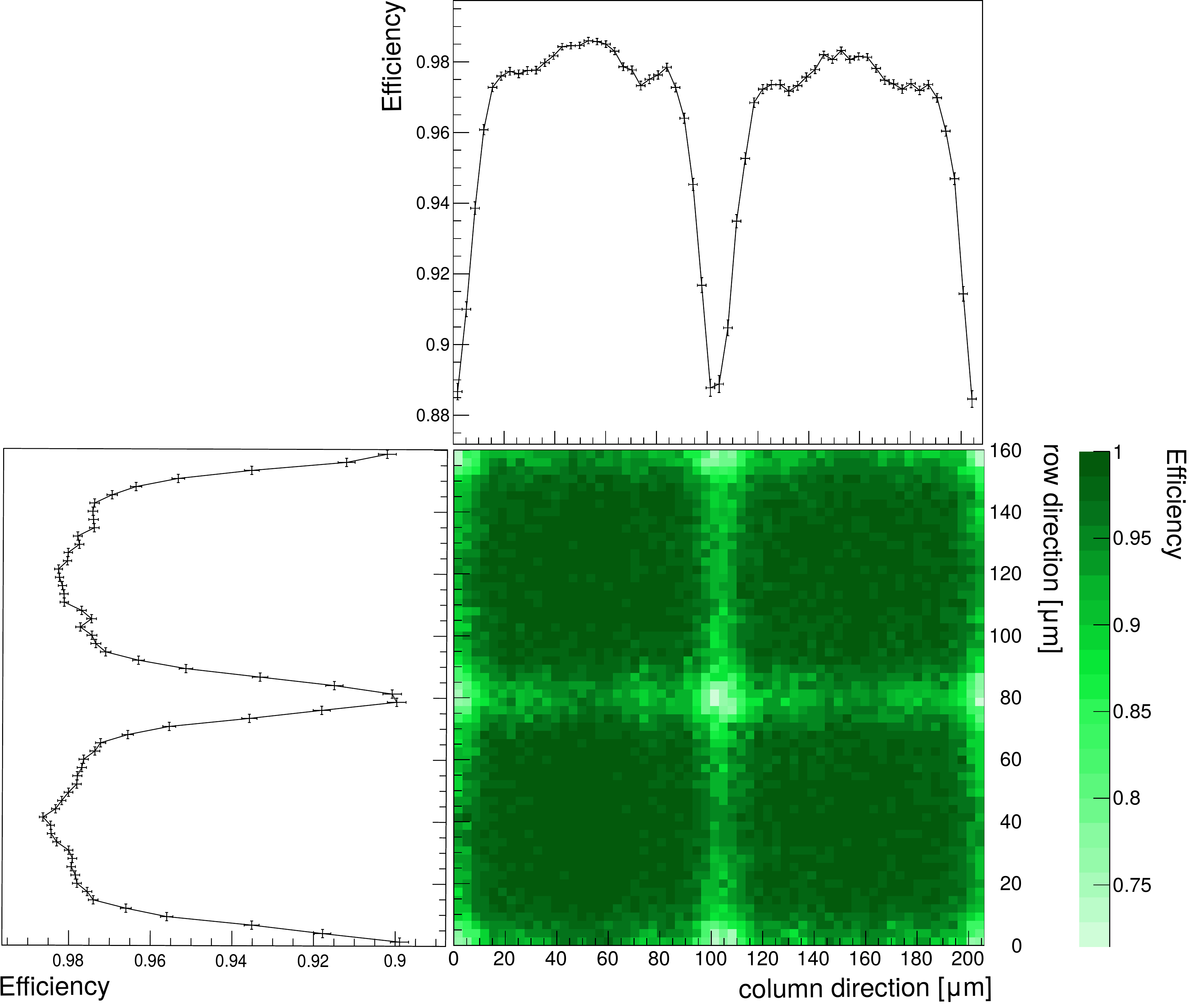}
  \caption{Efficiency, measured with HV=\SI{-40}{V} and a threshold of
    \SI{70}{mV}. Sub-matrices of 2$\times$2 pixels are stacked on top of each
    other, excluding the outer two columns / rows.}
  \label{fig:efficency_folded_40V}
\end{figure}

%
 
Straight tracks were fitted through the six telescope planes with an efficiency 
of $\sim$ \SI{80}{\%} (\SI{70}{\%}) with a DUT rotated by \SI{0}{\degree}
(\SI{45}{\degree}) in the beam. These tracks were then extrapolated to the DUT with an
expected position uncertainty on the DUT of \SI{4}{\micro m} (\SI{11}{\micro m}) for a
DUT  rotated by \SI{0}{\degree} (\SI{45}{\degree}). 
Readout events containing a single track of good quality 
and less than 50 hits on the \mupixS serve as 
the reference sample. If a hit within \SI{150}{\micro m} of the 
extrapolated track intersection with the \mupixS is found, track and hit are
matched. In addition, one noisy pixel on the \mupixS with a hit rate two orders 
of magnitude above the average pixel rate was excluded from the analysis.


We obtained data samples with two different configurations of the \mupixS: at the optimal 
working point with \SI{-85}{V} high voltage and a threshold of \SI{65}{mV}, 
and at  \SI{-40}{V} high voltage and a threshold of \SI{70}{mV} corresponding 
to lower efficiency settings. 
In both cases, a tune of the on-chip bias currents with a power 
consumption of \SI{300}{mW/cm\squared} was applied. 
For each individual pixel, the local threshold was adjusted to achieve a noise rate below \SI{1}{Hz}. 
With the first settings, an average efficiency of \SI{99.3}{\%} was measured 
(excluding the outer two columns / rows in order to avoid edge effects
and the non-connected diodes in the top row), see
the efficiency map in figure~\ref{fig:efficency_map_85V}.
Inefficiencies are visible mainly in the corners of the pixels, where the charge
is shared between four pixels, as can be clearly seen in 
figure~\ref{fig:efficency_folded_85V}, for which all cells of two by two pixels
(excluding again the outer two columns / rows) have been stacked on top of each
other in order to increase statistics.

In order to study the charge collection processes affecting the efficiency in 
more details, we analyzed the data taken with the reduced high voltage, resulting 
in \SI{96.1}{\%} overall efficiency; see figure~\ref{fig:efficency_map_40V} for an
efficiency map and figure \ref{fig:efficency_folded_40V} for the stacked two by two
pixel cell.
As expected, an efficiency drop is primarily observed along the pixel edges, where
charge is shared between two pixels.
It also becomes evident that hits in the central diode containing the amplifier
are more efficiently detected.

\section{Timing performance}

\begin{figure}[tb!]
	\centering
		\includegraphics[width=0.45\textwidth]{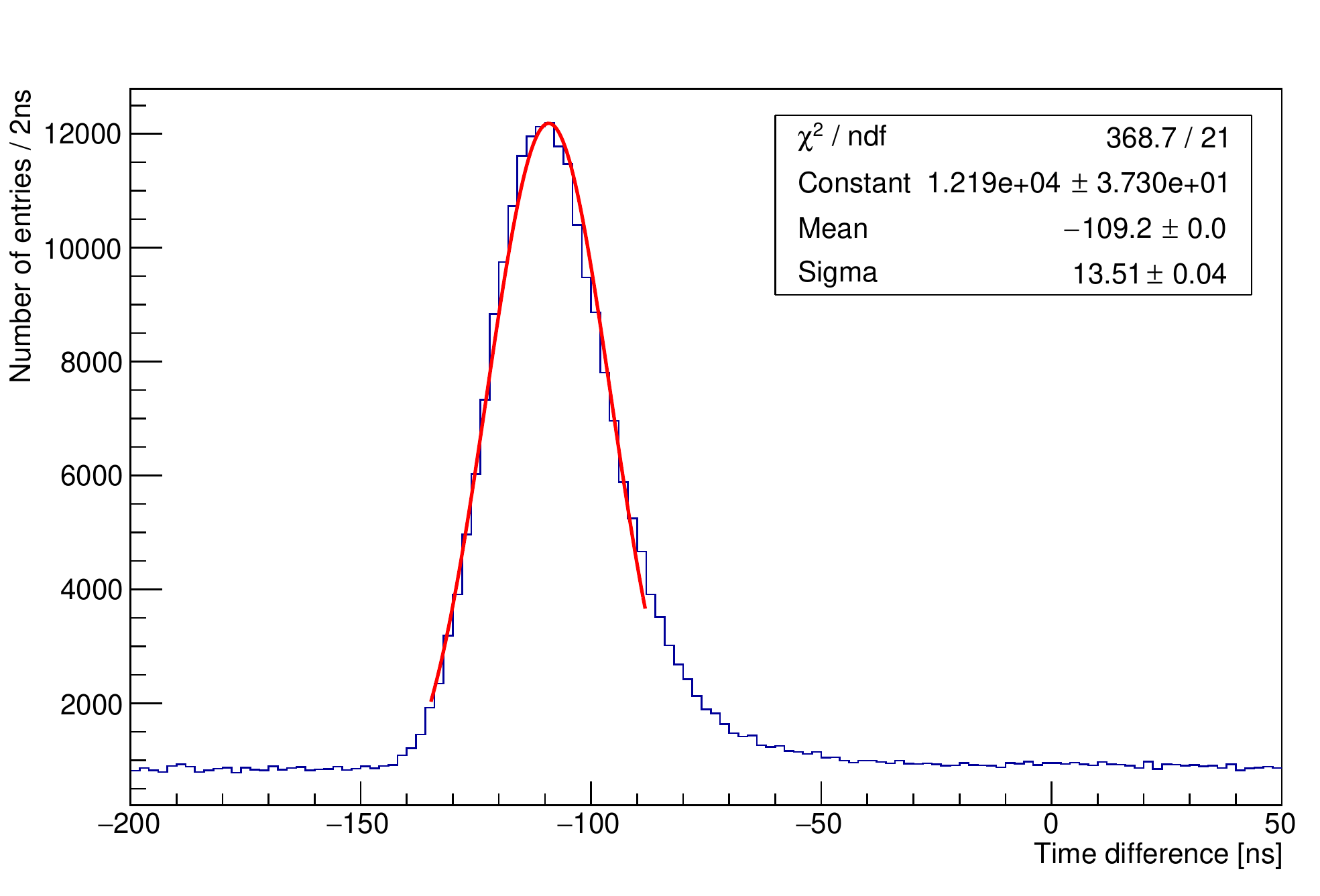}
	\caption{Difference between the hit timestamp and the scintillator coincidence time 
	for hits matched to a track measured with HV=\SI{-85}{V} and a threshold of \SI{65}{mV}.
	The offset of \SI{110}{ns} is given by differences in the cabling- and processing
	delays.
	A Gaussian distribution is fitted to the peak region of the difference distribution.}
	\label{fig:timediff}
\end{figure}

\begin{figure}[tb!]
	\centering
		\includegraphics[width=0.40\textwidth]{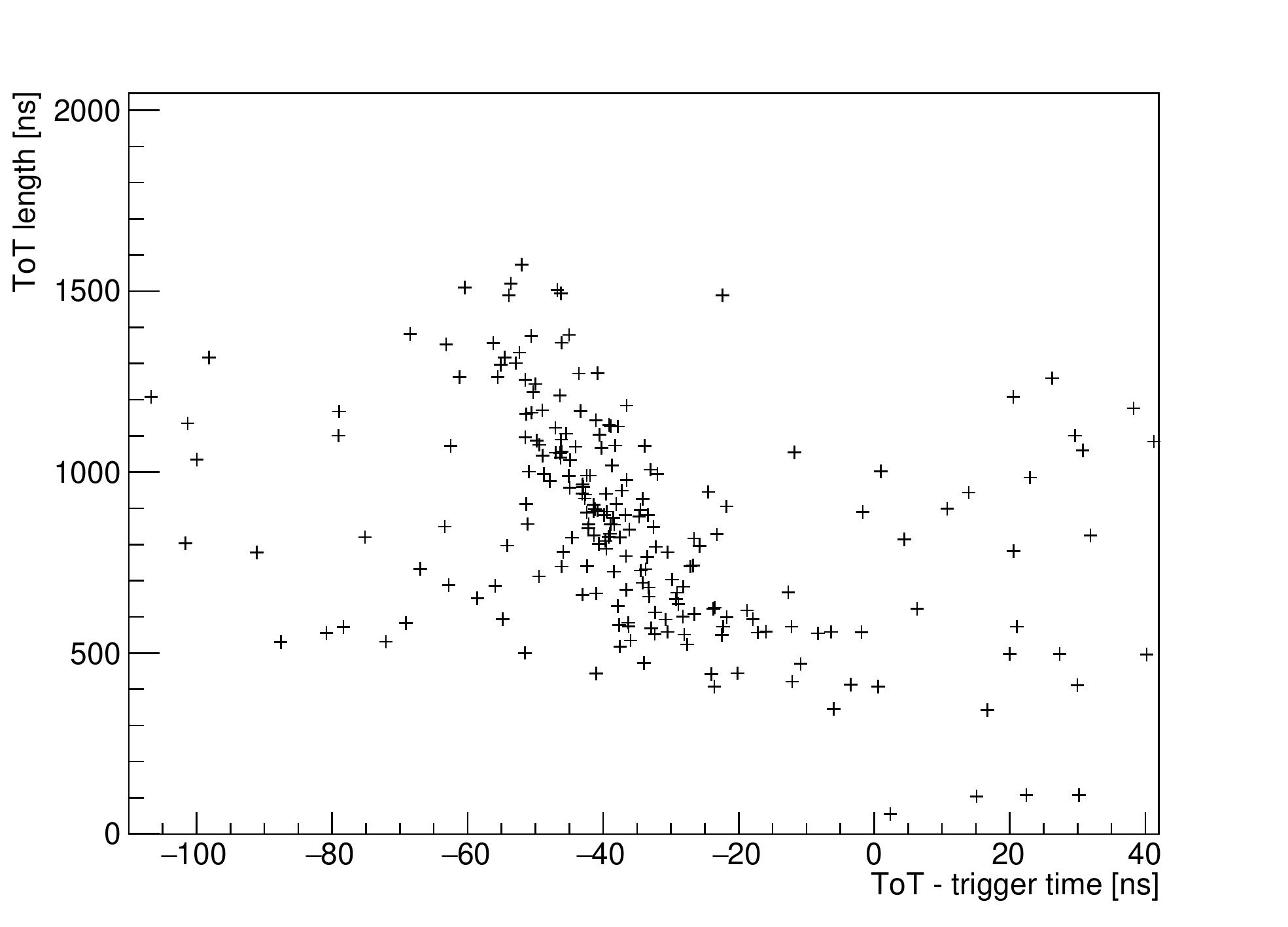}
	\caption{Time over threshold
          length versus the difference between the hit timestamp and the
          scintillator coincidence. Only hits matched to a track are shown,
          measured with HV=\SI{-85}{V} and a threshold of \SI{65}{mV}. The ToT
          information is only read for one pixel at a time, hence the
          reduced statistics.
	}
	\label{fig:tot_timediff}
\end{figure}

\FloatBarrier

\begin{figure*}[tp!]
  \centering
  \includegraphics[width=0.45\textwidth]{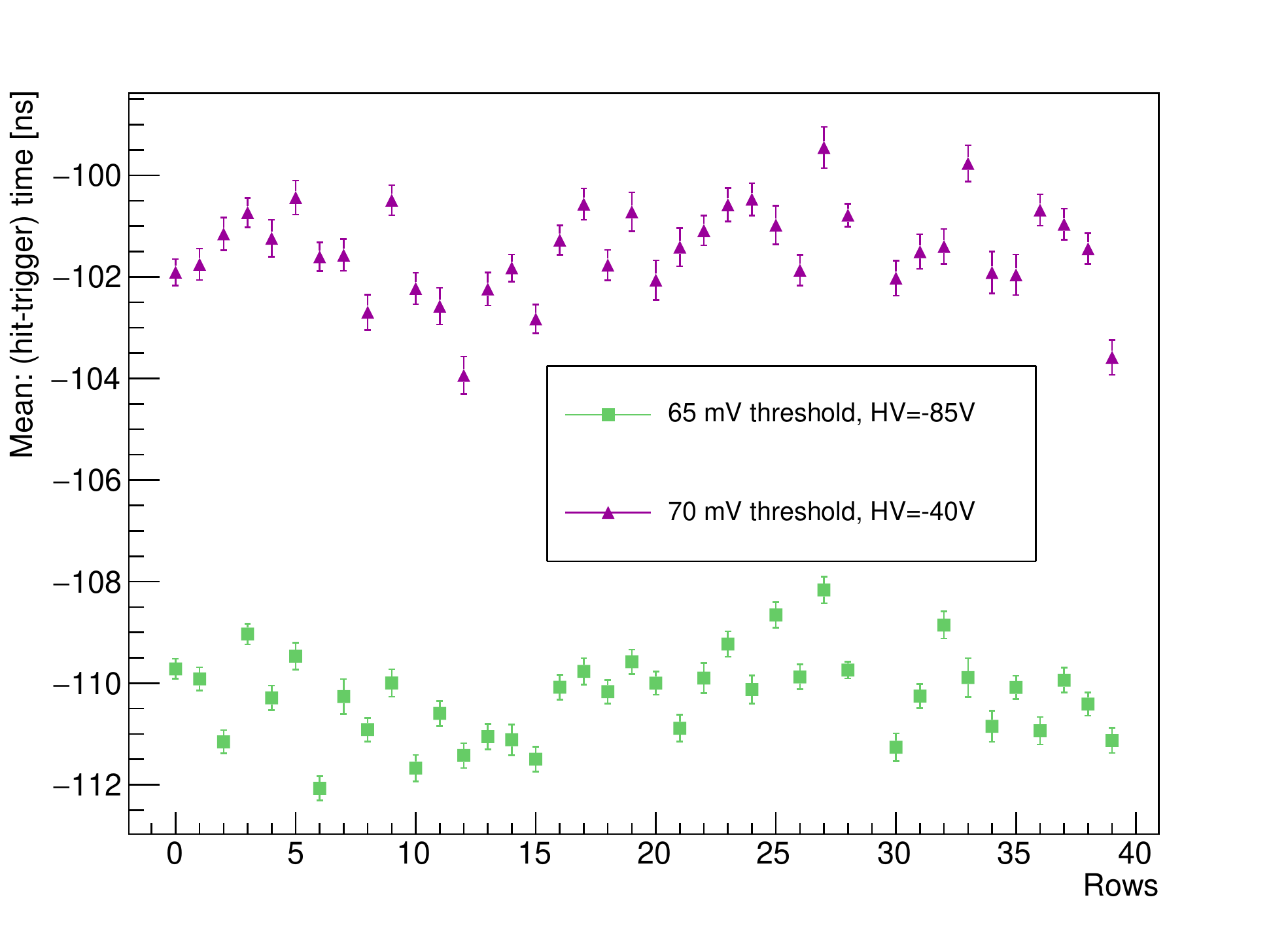}
	\includegraphics[width=0.45\textwidth]{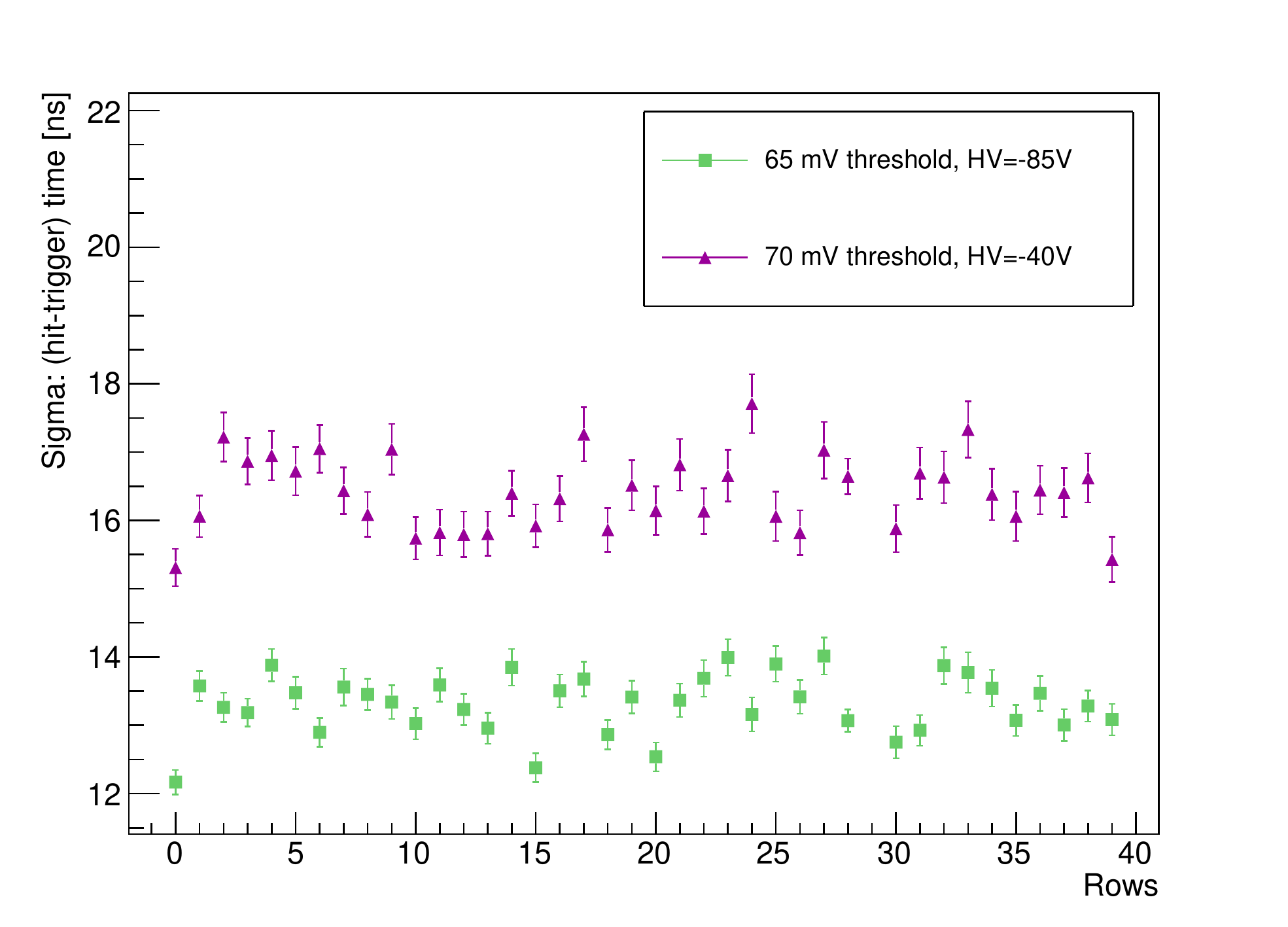}
  \caption{Mean (left) and width sigma (right) of the difference between hit and scintillator coincidence timestamp, determined from a Gaussian fit to the distribution for each row; measured with HV=\SI{-85}{V} and a threshold of \SI{65}{mV} (green squares) as well as HV=\SI{-40}{V} and a threshold of \SI{70}{mV} (purple triangles).}
  \label{fig:mean_timediff_row_40V}
\end{figure*}

\begin{figure*}[tp!]
  \centering
  \includegraphics[width=0.45\textwidth]{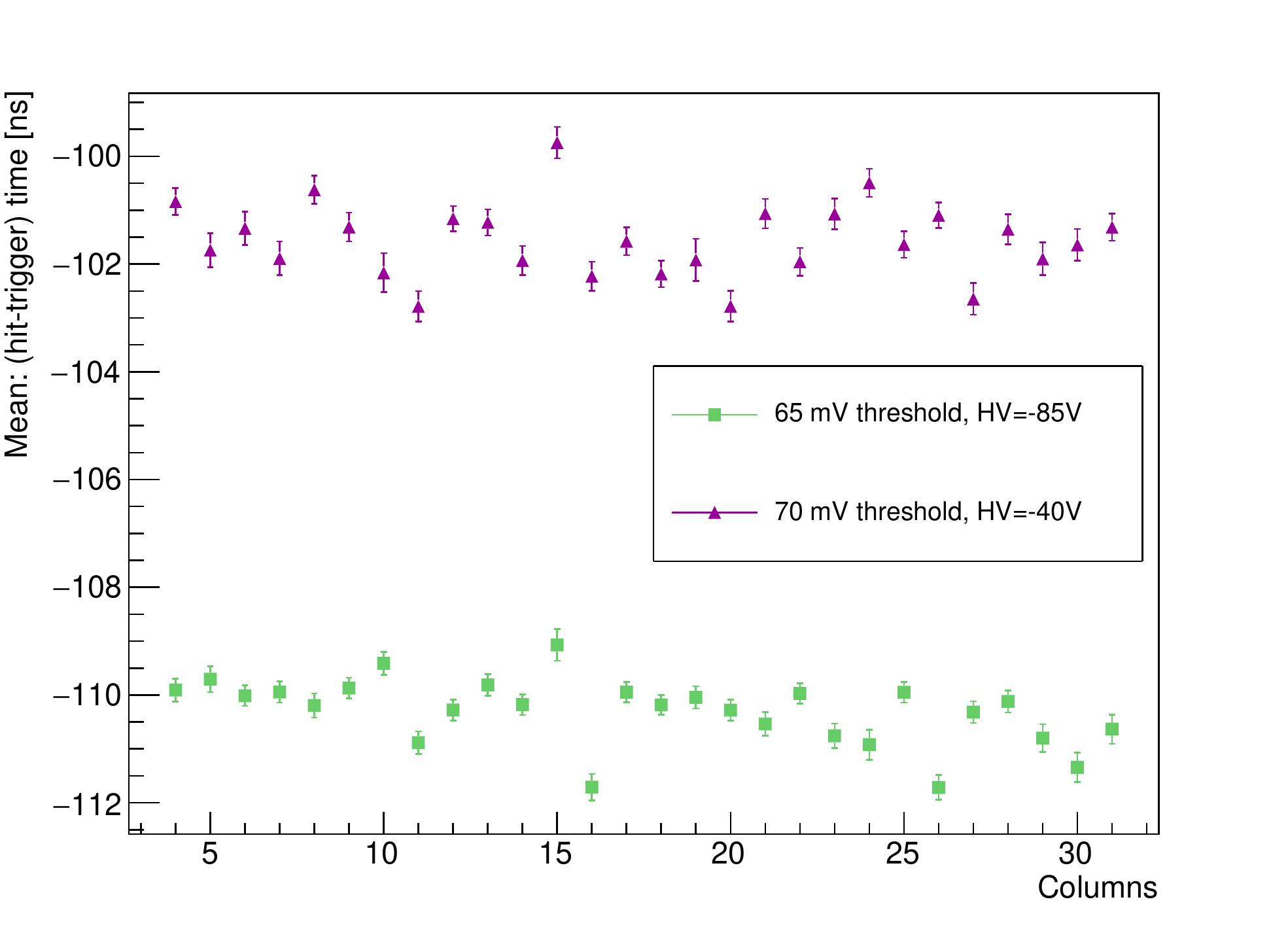}
	\includegraphics[width=0.45\textwidth]{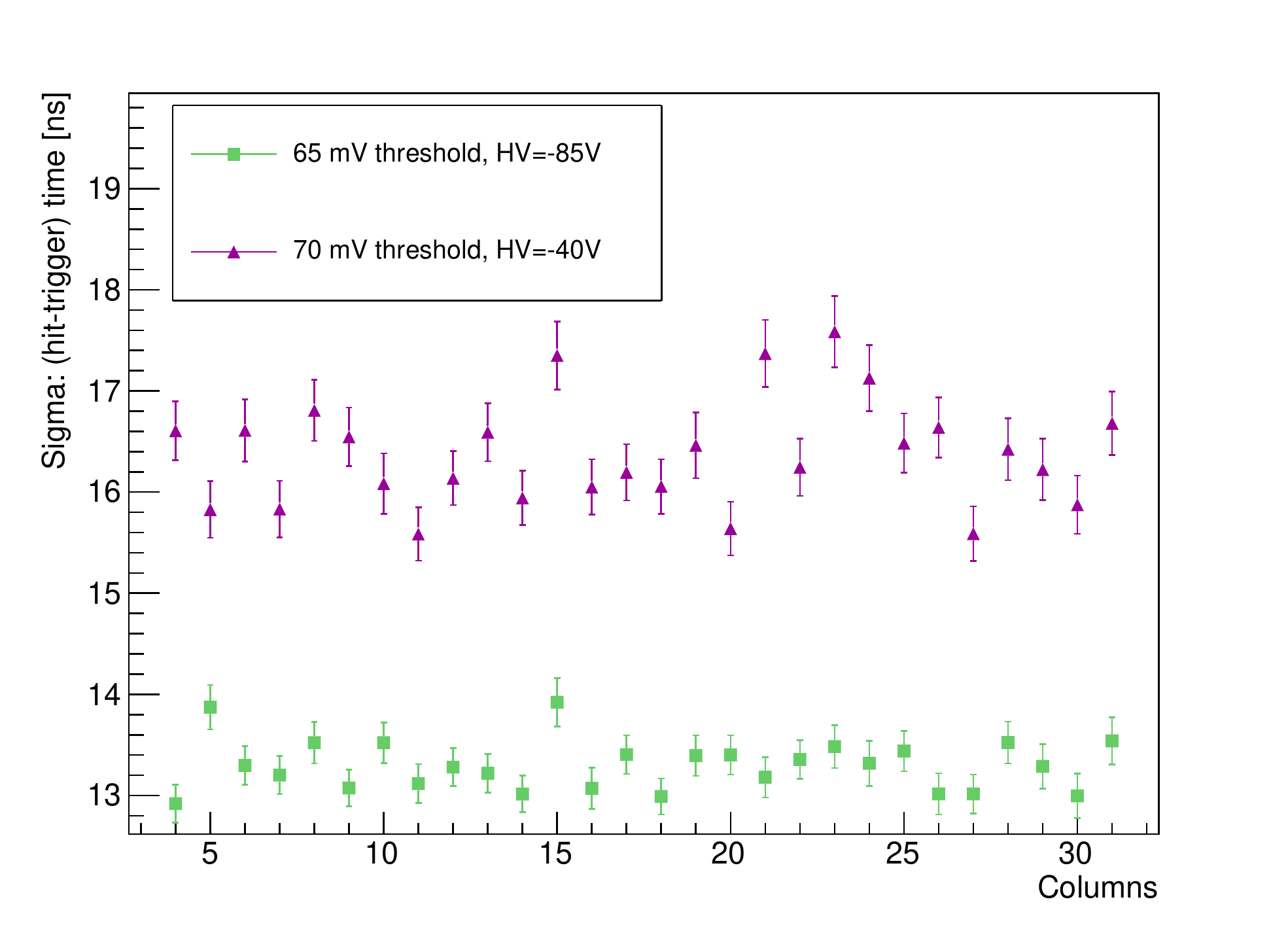}
  \caption{Mean (left) and width sigma (right) of the difference between hit and scintillator coincidence timestamp, determined from a Gaussian fit to the distribution for each column; measured with HV=\SI{-85}{V} and a threshold of \SI{65}{mV} (green squares) as well as HV=\SI{-40}{V} and a threshold of \SI{70}{mV} (purple triangles).}
  \label{fig:mean_timediff_col_40V}
\end{figure*}

\begin{figure*}[tp!]
  \centering
  \includegraphics[width=0.45\textwidth]{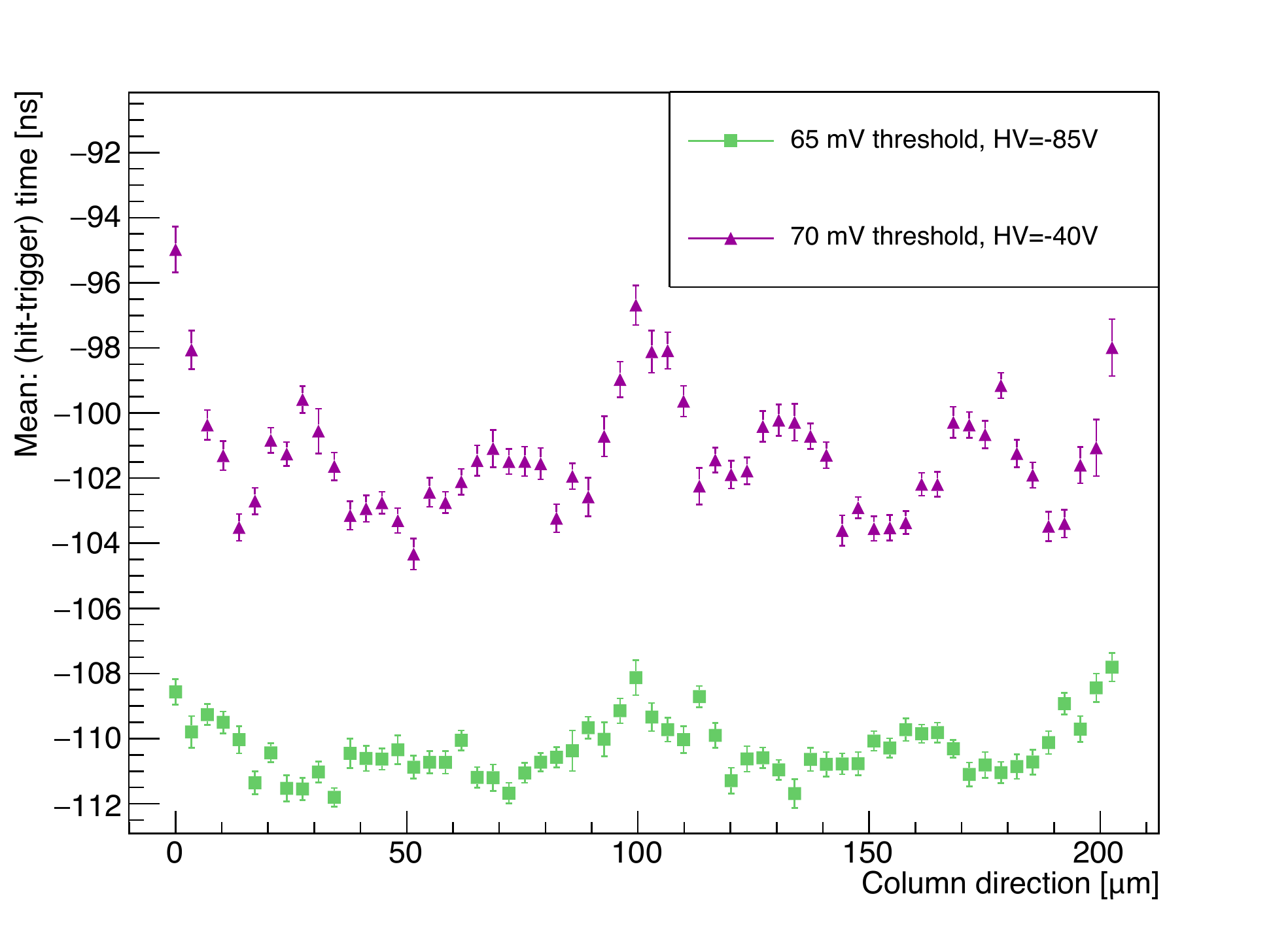}
	 \includegraphics[width=0.45\textwidth]{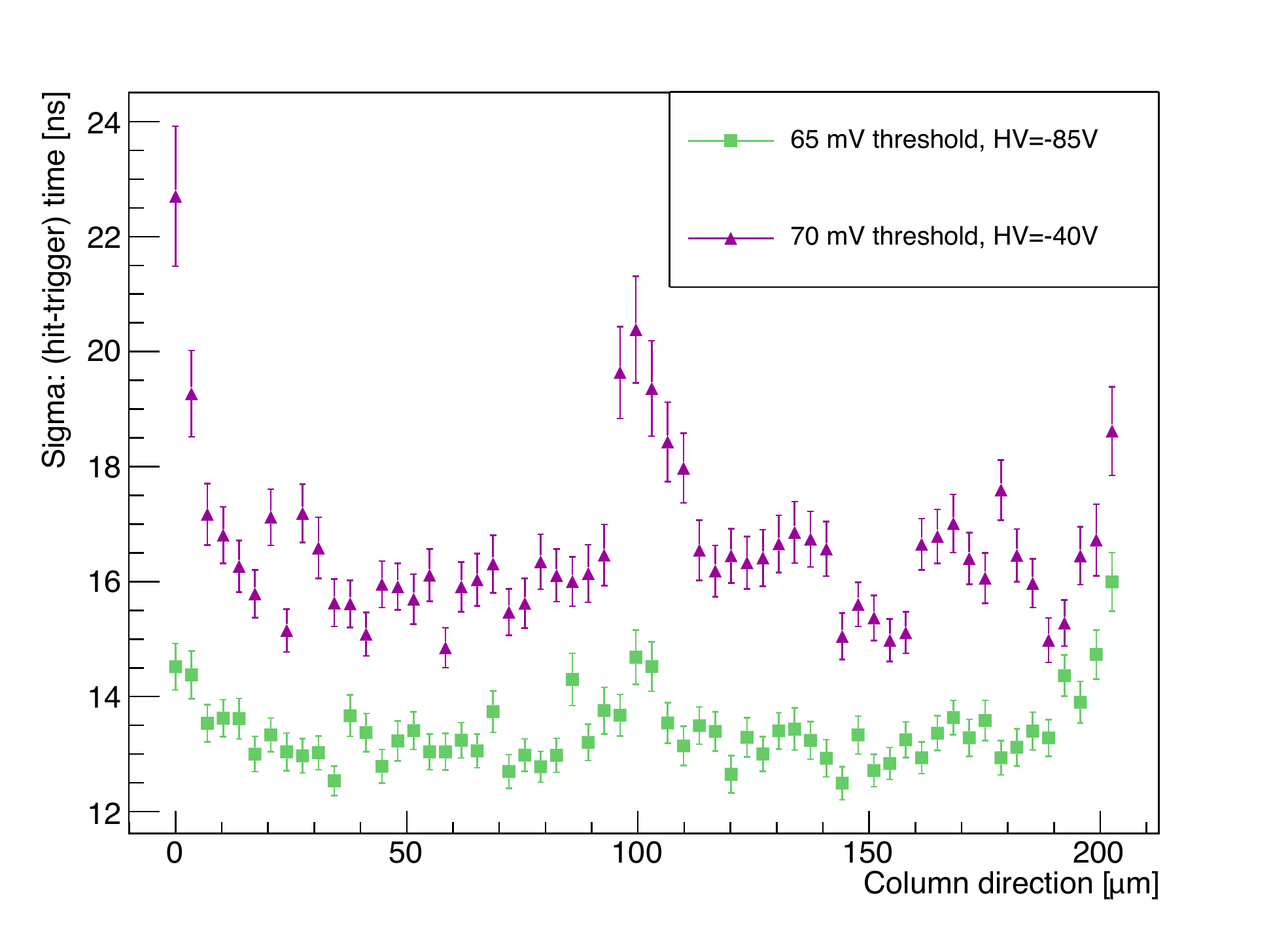}
  \caption{Mean (left) and width sigma (right) of the difference between hit and scintillator coincidence timestamp,
    determined from a Gaussian fit to the distribution for slices of
    \SI{3.4}{\micro m} width in column direction; measured with HV=\SI{-85}{V}
    and a threshold of \SI{65}{mV} (green squares) as well as HV=\SI{-40}{V} and a threshold
    of \SI{70}{mV} (purple triangles). Sub-units of 2 columns are stacked on top of
    each other.}
  \label{fig:mean_timediff_folded_col_40V}
\end{figure*}
 
\begin{figure*}[tp!]
  \centering
  \includegraphics[width=0.45\textwidth]{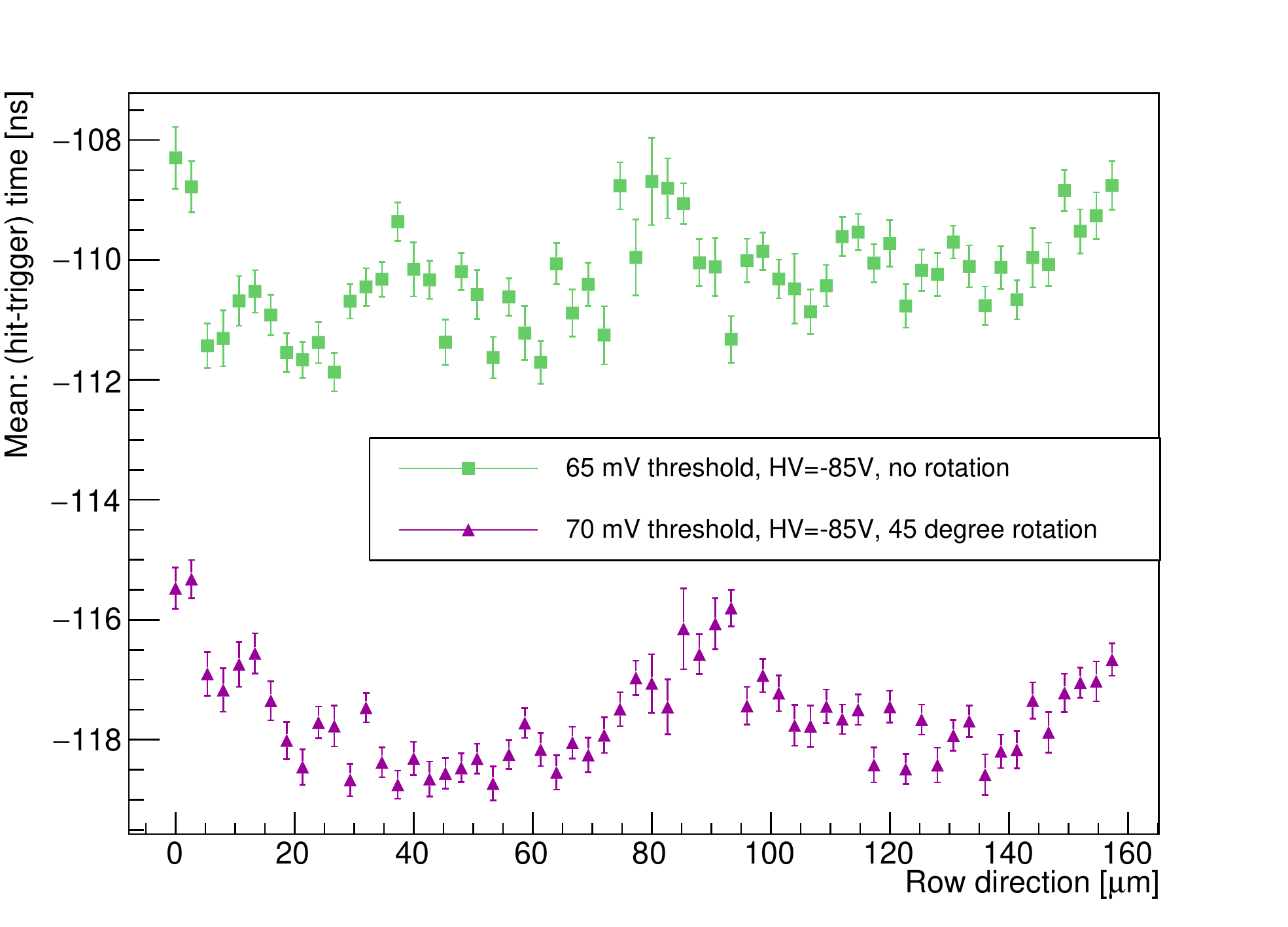}
	\includegraphics[width=0.45\textwidth]{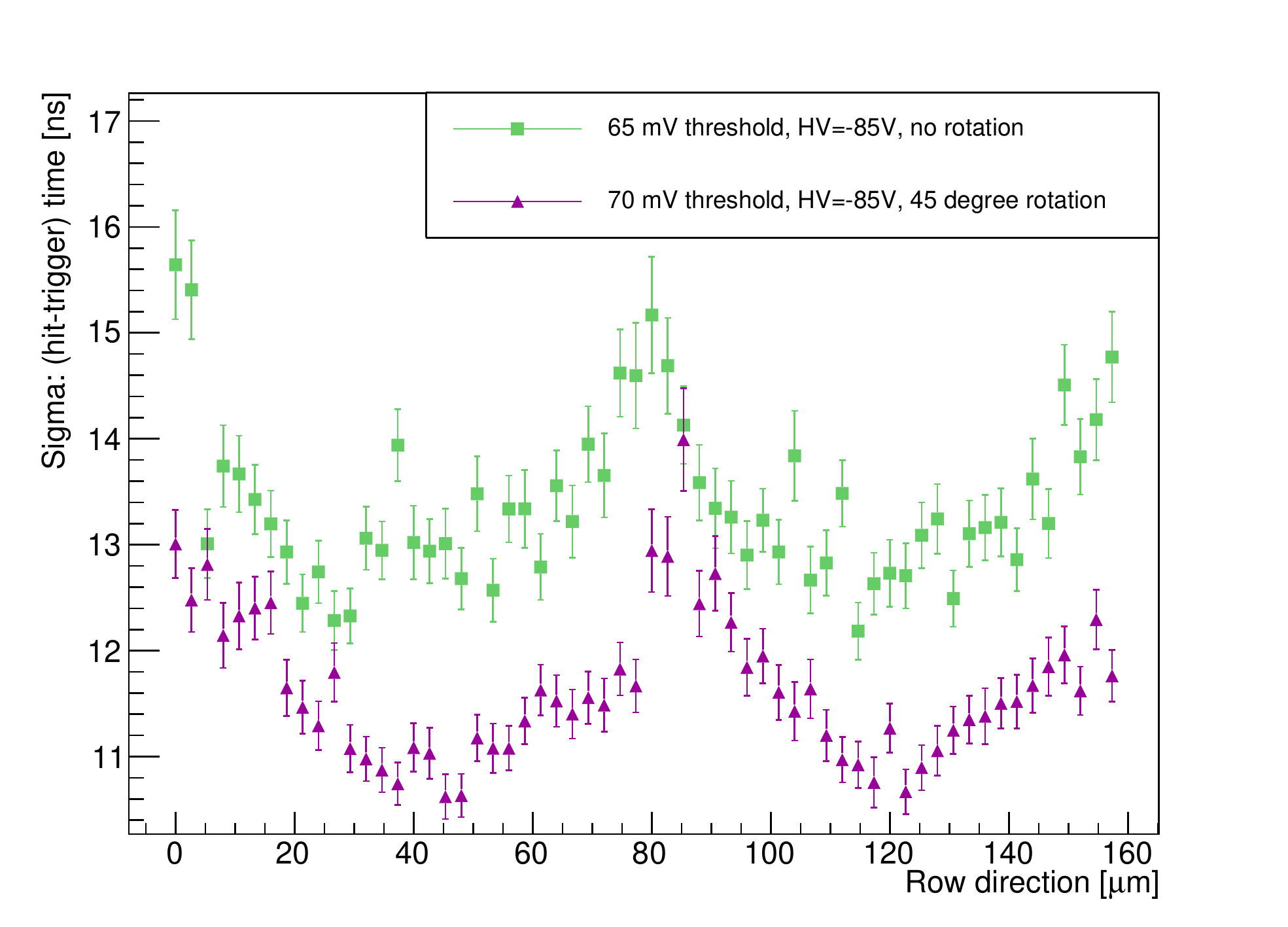}
  \caption{Mean (left) and width sigma (right) of the difference between hit
    and trigger timestamp, determined from a Gaussian fit for slices of
    \SI{2.7}{\micro m} width in row direction; measured with HV=\SI{-85}{V}, a
    threshold of \SI{70}{mV} and a rotation of 45 degrees (purple triangles) as well as
    HV=\SI{-85}{V}, a threshold of \SI{65}{mV} and no rotation (green squares), sub-units of 2
    rows are stacked on top of each other.}
  \label{fig:mean_sigma_timediff_row_rotation}
\end{figure*}

%
 %

The time resolution was studied by comparing the pixel hit times and the time
of the scintillator coincidence, see figure~\ref{fig:timediff}. 
A Gaussian function is fitted to the peak region of the 
distribution and a core 
time resolution of \SI{13.5}{ns} (with a negligible contribution from the 
scintillator setup) is obtained. 
There is however a sizable tail towards late
\mupix hits caused by the time-walk of small signals.
For one pixel per \mupixS, the time over threshold information is read out, it is plotted
in figure \ref{fig:tot_timediff}
versus the time difference between hits in this pixel and the scintillator
coincidence, clearly exhibiting the time-walk behaviour. 
The hit timestamp in the \mupixS is saved when the pixel signal crosses the
(fixed) comparator threshold.
The time resolution can thus be influenced by variations in the signal size
due to fluctuations in the collected charge, pixel-to-pixel production variations
in the amplifier and comparator circuits, drops of the high, low and bias 
voltages over the chip as well as the length of signal pathways. 

In order to study the spatial variation of the time resolution, 
we analyzed slices both in column- and row-direction separately. 
For the spatial distinction, the extrapolated telescope track position of the 
matched hit was used. 
For the selected hits, the differences between the hit and all recorded
scintillator timestamps of that event were studied using the mean (delay) 
and width (resolution) obtained from the Gaussian fit.  

Figures~\ref{fig:mean_timediff_row_40V} and \ref{fig:mean_timediff_col_40V} show 
the average delay and the resolution of the time measurement in dependence of
the row and column axis.
In particular the delays vary beyond the purely statistical expectation;
however no significant slope is apparent, which indicates that voltage drops 
over the small chip and signal transmission times do not significantly contribute.
Decreasing the high voltage and increasing the threshold leads to later hits with
a larger dispersion in time, but also here no structures on the chip are 
apparent.

In order to study the influence of the signal size on the time measurement, we
perform an average over all pixels and study delay and resolution as a function
of the hit position in the pixel; towards the edges and corners, charge is 
shared between several pixels and the signal seen by the pixel under study is 
thus smaller.
This effect is clearly visible in figure~\ref{fig:mean_timediff_folded_col_40V},
where we show the fitted delay and
width averaged over two columns using the reduced high voltage/efficiency settings.
The signal arrives later and the time resolution is worse at the pixel edge;
in addition, effects from the diode structure shown in 
figure~\ref{fig:MUPIX7_Sensor} become visible.
At \SI{-40}{V}, the region between the diodes is not completely depleted and 
part of the deposited charge is lost due to recombination or collected much slower
via diffusion. In addition, the electrical connections between the nine diodes 
are not ideal conductors; the effective detector capacitance thus becomes a 
function of the hit position within the pixel.


A comparison -- using the nominal high voltage and a low threshold --
 between an unrotated sensor and a sensor rotated horizontally by
45~degrees relative to the incoming particle direction (leading to a signal 
enhanced by a factor $\sqrt{2}$ compared to perpendicular operation) is
shown in figure~\ref{fig:mean_sigma_timediff_row_rotation}. The
projected pointing resolution in column direction is worsened by the rotation,
we thus show the row projections, in which the pixel edges are still visible 
but significantly reduced and the diode structure is no longer discernible for
the rotated sensor.
%
Not surprisingly, a large signal greatly improves the timing performance; in 
future versions of the \mupix we consequently plan to use a higher resistivity
substrate which will lead to a thicker depletion zone and thus a larger collected
charge. 

\section{Simulation}
\label{sec:Simulation}

\begin{figure}
	\centering
\includegraphics[width=0.49\textwidth]{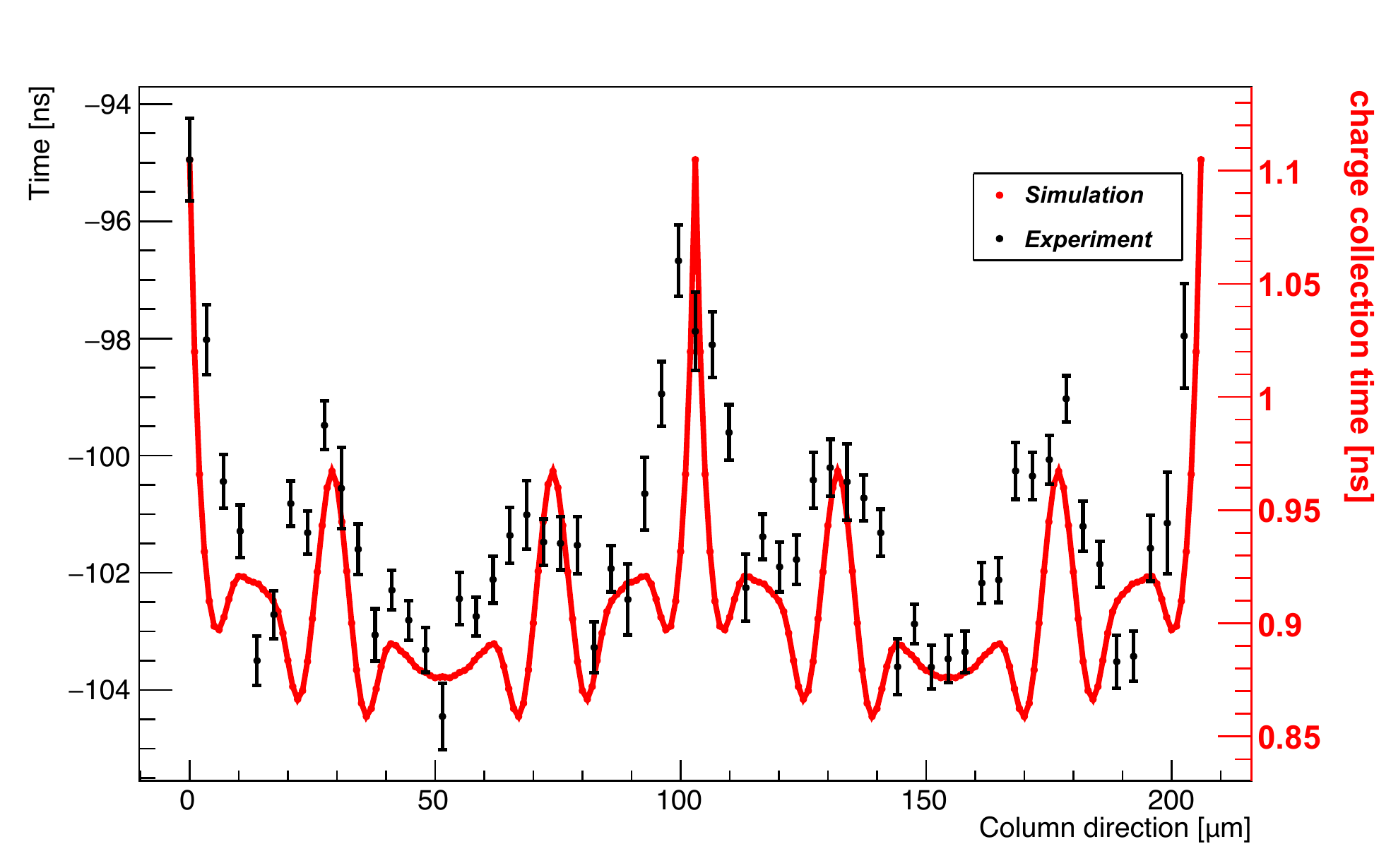}
	\caption{Comparison of the measured delays (stacked two-pixel cells, left scale, black points with error bars) 
	and the simulated charge collection times (right scale, red curve), for a 
	\SI{10}{\micro m} wide slice in
	column direction in the centre of the pixel (see figure~\ref{fig:MUPIX7_Sensor}) measured with HV=\SI{-40}{V}.
	Note that the simulation does not include the tracking resolution and the
	amplification and hit detection circuitry.}
	\label{fig:Simulation40V}
\end{figure}

\begin{figure}
	\centering
\includegraphics[width=0.49\textwidth]{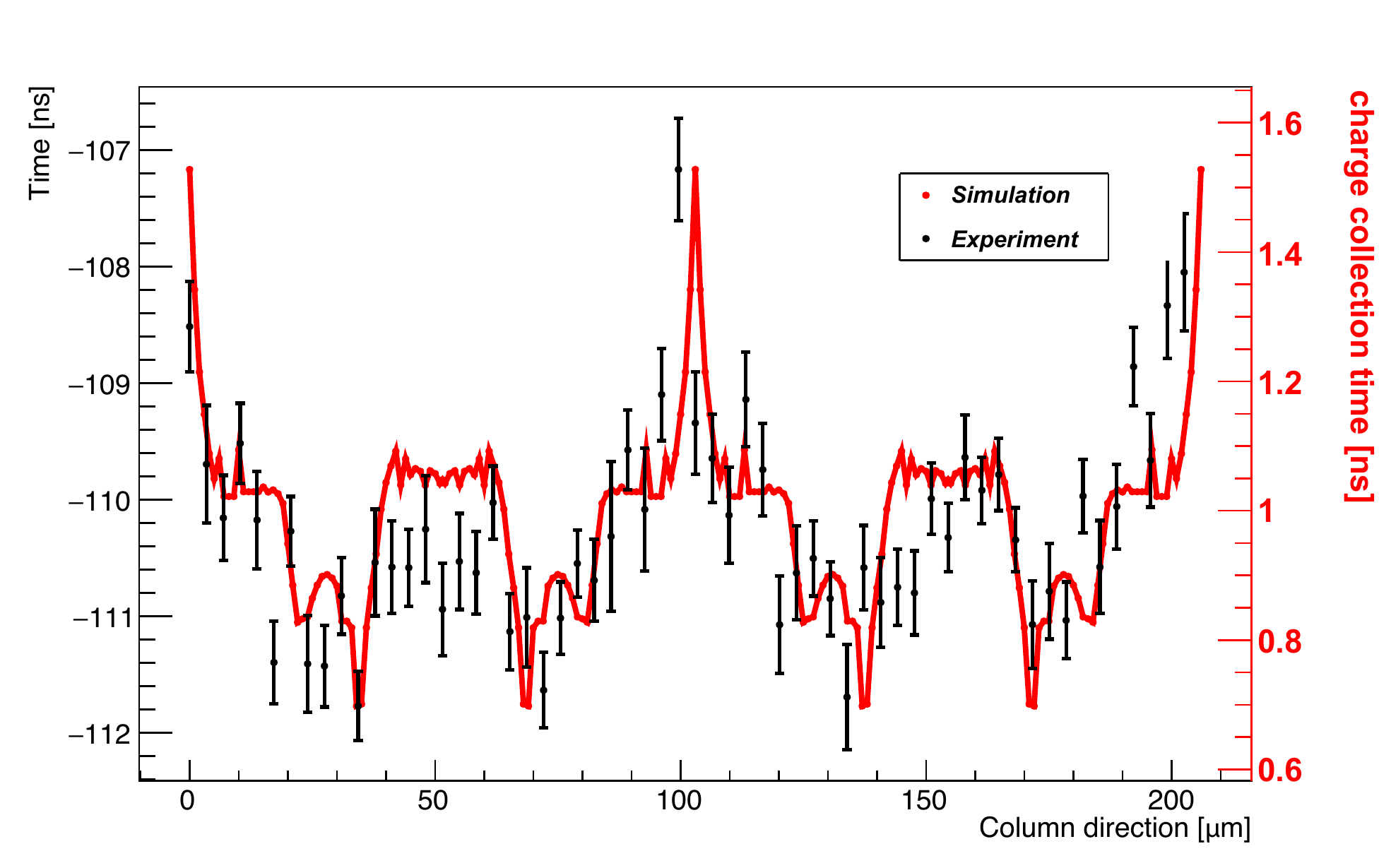}
	\caption{Comparison of the measured delays (stacked two-pixel cells, left scale, black points with error bars) 
	and the simulated charge collection times (right scale, red curve), for a 
	\SI{10}{\micro m} wide slice in
	column direction in the centre of the pixel (see figure~\ref{fig:MUPIX7_Sensor}) 
	measured with HV=\SI{-85}{V}.
	Note that the simulation does not include the tracking resolution and the
	amplification and hit detection circuitry.}
	\label{fig:Simulation85V}
\end{figure}

We have simulated the charge collection in the pixel using Synopsys TCAD 
(Technology Computer Aided Design) tools \cite{Synopsys}.
The 2-D simulation models a slice through the centre of the pixel in column 
direction (see figure~\ref{fig:MUPIX7_Sensor}). A linear energy transfer along the particle track of 
\SI{20}{\atto C \per \micro m} was assumed in the heavy ion model of 
Sentaurus device \cite{Sentaurus}. Particle incidence is at a right angle to the sensor and the
impact point is varied along the column direction. The charge collection curve
in the diode at \SI{-40}{V} and \SI{-85}{V} is fitted with an exponential and 
the time constant is determined.  The charge collection time is expected to be linearly
correlated to the measured latency by an unknown scale factor determined by the
amplifier and comparator circuitry. The simulated time constants are plotted in 
figures~\ref{fig:Simulation40V} and \ref{fig:Simulation85V} and compared to
the measurements, selecting a \SI{10}{\micro m} wide slice in the centre of the
pixel (figure~\ref{fig:MUPIX7_Sensor}). Taking into account the limited position 
resolution of the measurement, all major features are reproduced by the simulation, 
giving us confidence that simulations can be used to further optimize the charge 
collection and timing behaviour in future devices.
 
\section{Conclusions}

Using the test beam facility at DESY and an EUDET pixel telescope we have
performed studies of efficiency and timing resolution of the \mupixS 
high voltage monolithic active pixel sensor.
At the nominal operation point, the sensor reaches an efficiency of \SI{99.3}{\percent}
for perpendicular \SI{4}{GeV} electron tracks.
The remaining inefficiency is due to charge sharing between pixels, particularly
in the pixel corners; no effect of the segmentation of each individual pixel 
into nine electrodes is visible.

The time resolution at nominal settings is about \SI{14}{ns}; our study shows 
that it is dominated by time-walk, which can be easily seen at the
pixel edges, where the signal size in the pixel is reduced due to 
charge sharing between pixels.
Measurements with reduced efficiency settings make in-pixel variations due to
the split charge collection diode visible and are well modeled by a TCAD 
simulation.
With the larger charge deposited by particles crossing at 45~degrees to the
sensor planes, both inefficiencies and sub-pixel timing variations are hardly 
observable.
For future versions of the \mupix, we thus intend to use a higher resistivity 
substrate, leading to an increased collected charge. We will also explore several
schemes for time-walk compensation with the aim of a further reduction of the
time resolution.
 

\section*{Acknowledgments}

The support of the Deutsches Elektronensynchrotron (DESY, Hamburg), a member of the 
Helmholtz Association (HGF) providing the test 
beam and the related infrastructure made this measurement possible. We would 
especially like to thank the EUDET telescope group at DESY for their valuable
support of this test beam measurement.

A.~Meneses Gonzales and H.~Augustin would like to thank Marta Baselga of the 
Institut f\"ur Experimentelle Teilchenphysik (ETP) des Karlsruher Instituts 
f\"ur Technologie (KIT) for the comprehensive introduction to Synopsys TCAD.

N.~Berger, D.~vom Bruch, Q.~Huang, A.~Kozlinskiy and F.~Wauters would like to thank the Deutsche
Forschungsgemeinschaft for support through an Emmy Noether
grant and the PRISMA cluster of excellence at Johannes Gutenberg University Mainz.

M.~Kiehn, L.~Huth and S. Dittmeier acknowledge support by the International Max Planck
Research School for Precision Tests of Fundamental symmetries. 
H.~Augustin and A.~Meneses Gonzales acknowledge support by the HighRR research 
training group (GRK 2058) and A.K.~Perrevoort by the Particle Physics beyond the 
Standard Model research training group (GRK 1940).

\end{document}